# About a Discrete Cellular Soliton (computer simulation)

A. Kornyushkin

www.kornju.narod.ru, kornju@mail.ru

For the first time a mathematical object is presented - a reversible cellular Automaton - with many paradoxical qualities, the main ones among them are: a frequent quickly return to its original state, the presence of a large number of conservation laws and paradoxical "fuzzy" symmetries, which connects the current position of the automaton with its signature Main Integral.

## 1 Introduction

In 1980, the journal Science published an article of Feigenbaum M. J. **Universal Behavior in Nonlinear Systems** [1]. This article was a simple mathematical observation concerning one-dimensional maps. Of particular interest in this situation was the fact that the observation was made by simply using a calculator! And over time this work has become a progenitor of serious mathematical sciences.

Now, in comparison with 1980, the situation has changed dramatically. Home computers have appeared. This allows us to do much more serious and profound observations. We are presenting one of them. We called it the phenomenon of Discrete Cellular Soliton (DCS). It refers to the reversible cellular automata, and so first let's talk what it is.

## 2 What is cellular automaton and reversible cellular automaton?

Let`s take the definition of **cellular automaton** from the Wikipedia.

A cellular automaton (pl. cellular automata, abbrev. CA) is a discrete model studied in computability theory, mathematics, physics, complexity science, theoretical biology and microstructure modeling. It consists of a regular grid of cells, each in one of a finite number of states, such as "On" and "Off" (in contrast to a coupled map lattice). The grid can be in any finite number of dimensions. For each cell, a set of cells called its neighborhood (usually including the cell itself) is defined relative to the specified cell. For example, the neighborhood of a cell might be defined as the set of cells a distance of 2 or less from the cell. An initial state (time t=0) is selected by assigning a state for each cell. A new generation is created (advancing t by 1), according to some fixed rule (generally, a mathematical function) that determines the new state of each cell in terms of the current state of the cell and the states of the cells in its neighborhood. For example, the rule might be that the cell is "On" in the next generation if exactly two of the cells in the neighborhood are "On" in the current generation, otherwise the cell is "Off" in the next generation. Typically, the rule for updating the state of cells is the same for each cell and does not change over time, and is applied to the whole grid simultaneously, though exceptions are known.

Everything is clear. But we introduce our little pre-determination. Let`s call "the set of cells, called *neighborhood*" - **the mask** of a cellular automaton. Let`s cover the mask with a minimum square. Let`s call his side - **k**. Then let`s call the quantity **[(k-1)/2]** – **a rank** of the mask. And then we give an example, taken from the same source - from the Wikipedia. The Game of Life.

The universe of the Game of Life is an infinite two-dimensional orthogonal grid of square *cells*, each of which is in one of two possible states, *alive* or *dead*. Every cell interacts with its eight *neighbours*, which are the cells that are horizontally, vertically, or diagonally adjacent. At each step in time, the following transitions occur: 1. Any live cell with fewer than two live neighbours dies, as if caused by under-population. 2. Any live cell with two or three live neighbours lives on to the next generation. 3. Any live cell with more than three live neighbours dies, as if by overcrowding. 4, Any dead cell with exactly three live neighbours becomes a live cell, as if by reproduction.

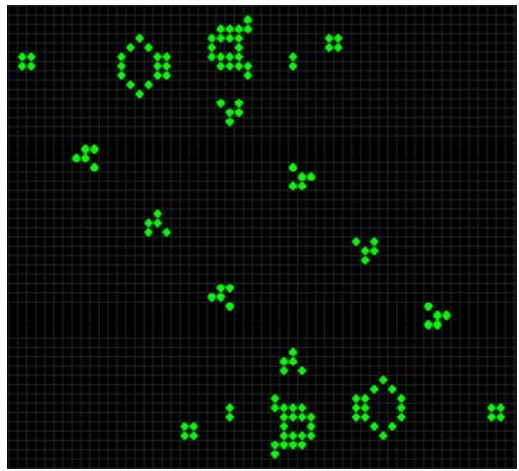

Fig.1. «Shootout» in the game of Life.

Then in the Wikipedia it is noted that these simple rules lead «to a huge variety of shapes that can occur in the game».

It is practically a common quality of many CA! Despite the fact that the laws of the transition and the initial conditions can be very simple, the behavior of the CA can be extremely difficult and unusual. So, it all takes place in today's science as "Cognition of Difficult things" and there is no, even the similarity of the general theory.

But we will talk about a certain subclass of the CA about which «at least something is known». In the book of Margopolus Toffoli [2] they are called as reversible cellular automata (RCA) of the second order. Since there is a "second" order, then there must be a "first". But we'll talk exactly about such RCA, and in our own signs.

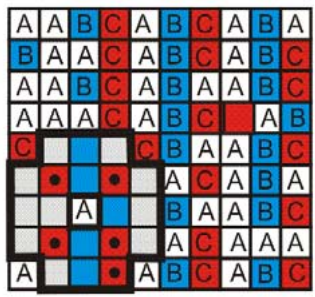

Fig. 2. The four main principles of reversible cellular automata devices of the second order.

This RCA has three state of the cell, and its structure is exhaustively illustrated in figure 2. We shall remember well that the cells **A** that we have are «white», the cells **B** are «blue», and the cells **C** are the «red»!

What is "known in advance" about it and why is it interesting? It is interesting that if you make the so-called **transliteration** - that is you replace all the cells **B** with **C** and all cells **C** with **B**, and, further, apply the same rules of transition... then the automaton **will go back in time**! (See [2], though it is obviously).

And if it is so, it means that any state of the cellular automaton always has exactly one previous, and it can be found easily! (To make a transliteration, and then to make a move back).

And if it is so, it means that such a cellular automaton will always go on its own Cycle, not overlapping with others, and if the total number of states is finite, **by all means it will pass through its starting point some time.**

After what period of time will the cellular automaton do it? Usually, if the Automaton does not have an obvious symmetry, it will do it after a very, very long period of time, the order of the total number of states: 3 in the degree of the total number of cells. And if the size of the area is 100 x 100, so it is number 3 in the degree of 100x100 = 10000. It is transcendentally great! Therefore, we will call a Cycle with such a number of steps a **Giant** Cycle.

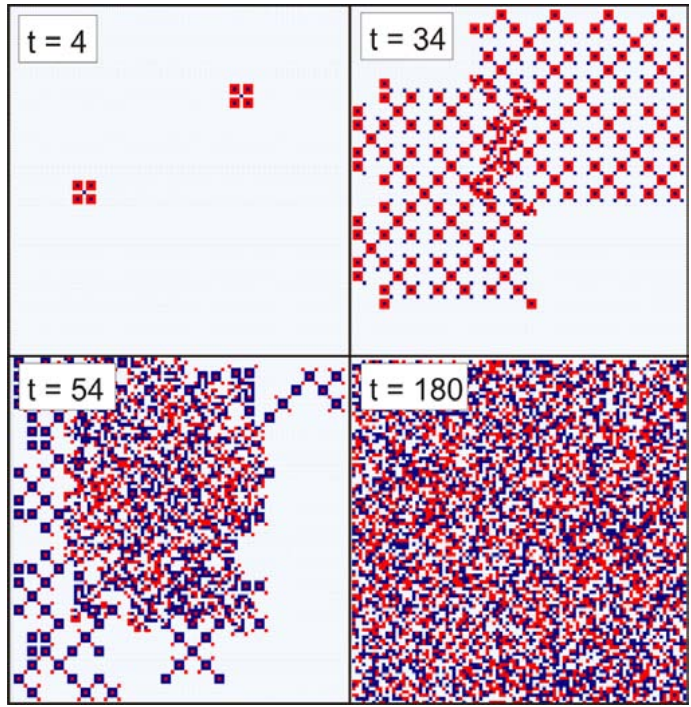

Fig. 3. An example of a typical movement of the RCA.

Figure 3 shows the typical movement of a random RCA. The way how Automaton goes to the Global Cycle. (His law is not given, and it is not necessary!) We started with two cells **B**, the rest - **A**, and after only 180 steps we got seething mess. One can say that we have visualized, what it means - number 3 in the degree of 10000.

And from the reason of simple common sense it follows that in general ANY reversible cellular Automaton should behave exactly that way! And from this common rigorous rule, nevertheless, there is EXACTLY ONE exception. And let's describe it.

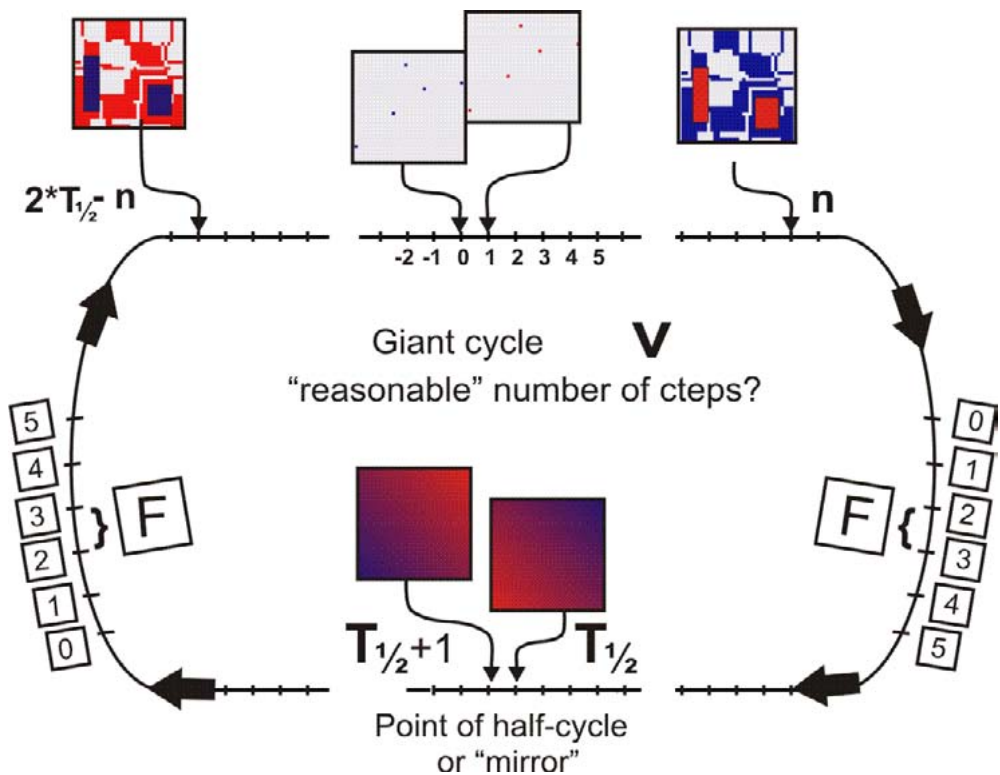

Fig. 4. What is the movement of the reversible cellular automaton (and our soliton, among others) with the beginning at white and blue plate.

So, we construct our cellular soliton. It is constructed extremely easy …

1). The laws of transition are as follows: if **any mask THAT HAS THE SYMMETRIES OF THE SQUARE has at least one point C - the law of transition is (II). If not - the law of transition is (I).**

Choose 10 – from the head! - such masks, the rank doesn`t exceed 4.

2). We choose the size of our square area as 70x70 - and we surely **CLOSE IT IN THE TORUS!**

3). Now it is left to specify the question of initial conditions…

We choose the following variant: all the cells are in the state **A** and only a few - random – are in the state **B** (Fig. 4). Then in the next step the blue ones will go into red, and the cells **A** will remain cells **A**... And this means that the automaton already at the first step **will pass into its transliteration**, and **immediately** will go back in time. And every state of the automaton at point of time **t** will be a transliteration of the automaton in the point of time - **(t-1)**. Some time these two movements - forward and back – will close up.

We call the moment when it happens as **a Point of «mirror»** or **a Point of half-cycle ($T_{1/2}$)**.

We choose and write down for further use 5 sets of 8 random cells **B**. («Random» - in the sense that the coordinates of cells are given by the library function rand (). X = rand ()% 70. Y = rand ()% 70. We want to recall that. All the other cells are cells **A**).

So in total we have 5 (the number of sets of initial points) * 10 (the number of masks) = 50 different cellular automata.

Let`s start them up…

## 3  First meeting with the Cellular Soliton

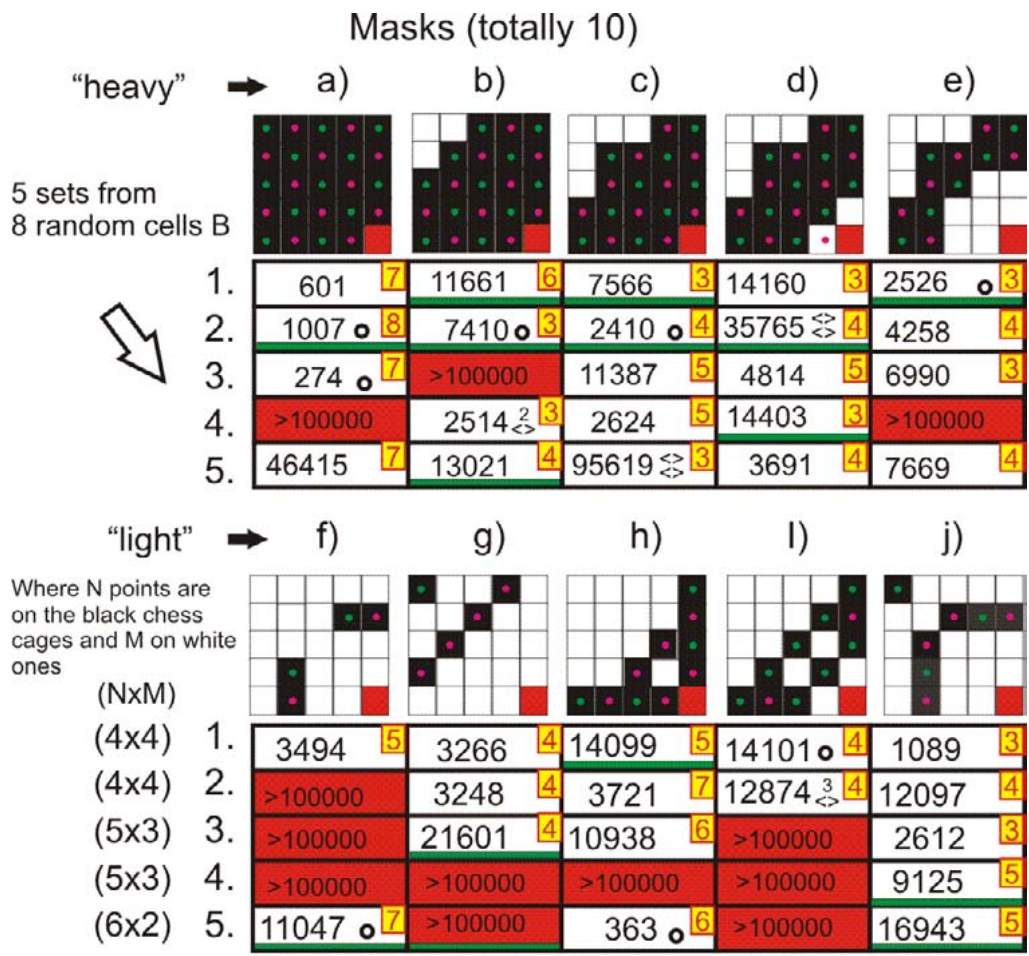

Fig. 5. The results of our tests. We note a surprising fact. The time of coming back doesn`t depend very much on the number of cells in the mask. In any case, the mask (j), in this sense, is one of the best.

The results are shown on the figure 5. On the top there are shown «quarters» of our masks (remember that they are symmetrical!), as far as we have imagination to invent them. At the intersection of the number of a set of 8 cells **B** on the horizontal line and masks on the vertical line, there is time when our cellular automaton reaches the Point of half-cycle. It is seen that in 38 of 50 cases this time is quite reasonable and a small number, and we are not talking about any 3 in degree of 4900 steps! This is our effect of CELLULAR SOLITON!

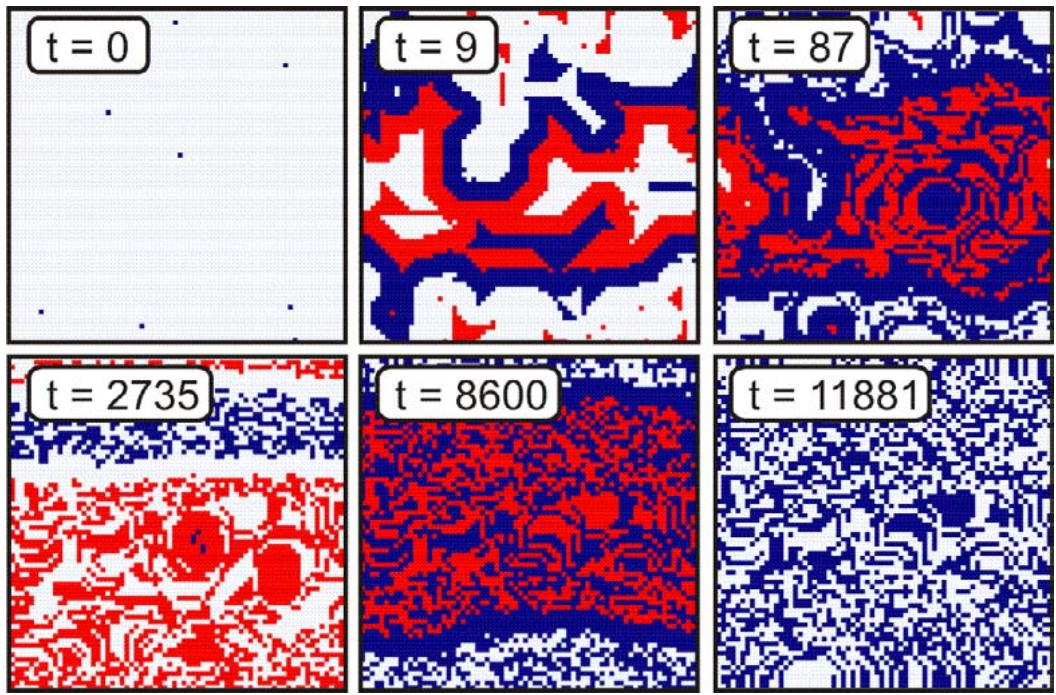

Fig. 6. Six frames from a motion of our soliton 1(b). At the very beginning – there are few blue ones. In the end (the Point of half-cycle) - about half of them are blue. All our solitons return exactly through such blue-and-white pictures.

Let's now look at the movement of any of them. Let it be a soliton 1(b). Six frames of this movement are shown on the figure 6. The frame t = 0 is the starting position. The frame t = 11881 – the point of mirror-half-cycle. So what happens?

Initially, just the first 10 - 20 steps, there is an "explosion". Rapid "erratic" movement, which then ... at some point of time "calm down" somehow and the automaton begins to "work with a period 6". (Period - is a number of steps through which our automaton comes in approximately the same condition. By the way, this gives us an opportunity to shoot cartoons.) Automaton distributes somehow its cells in 6 phases, and itself creates a very interesting structure - so-called **River System with Different Banks.**

In six steps there is no change at the Bank. And the Rivers, by contrast, are constantly twisting a little and changing their riverbeds. And just in place of these "twisting places" they are permanently as if "washing out and modifying the" the Bank at this place.

For simplicity let's introduce the following definitions. We call the River that encircles the whole torus - **SuperRiver**. One that doesn't encircle will be called just the **River**. (There are also very distinctive **NullRivers**, to which we will come later!)

In the Bank the most cells **CBCBAA** vary with a period 6 and with the appropriate phase shift. In the **Rivers** the cells vary with the period 3: **CBACBA**...

So...There has been an "explosion"! The Rivers were formed. And they were becoming stronger. That river which is bigger "eats" that river, which is smaller. To do this, the Rivers are sure to have to be in the appropriate phases. But fortunately the Rivers pulse, and at the time of passage through the zero they shift their phase. When the phases go well - the fusion takes place. They merge, merge...And our last SuperRiver formes. At some point in time it also pulses, but...

At this point, when passing through zero, it will turn out that on some of the phases all points **C** disappeared, and this means that the Automaton has reached a **Point of mirror**. We'll say that, in this case, the automaton **has closed**. (Or **has returned**). We have described a "canonical case". But if in the table in figure 4 there is a bold circle, it means that the automaton closed in a different way. Without using the last SuperRiver.

But, in general, all cases are of extremely peculiarity - we will talk more about it! (If in the table there is an icon ◇, and above it there is a figure **k** - this means that the last SuperRiver while pulsing "didn't close gently, through zero", but broke during the transition into **k** clusters which worked for some time before the final closure. If there are two icons ◇ - this means that once two SuperRivers were formed, which later for a long time tried to figure out which one of them is "more important". And so on ...)

We put our mask on a chess board and count how many cells of the mask fell on black squares, and how many of them fell on the white. It is easy to see that only masks, in which that number is approximately equal, work well. (The masks made up exclusively of black or exclusively of white cells - are "not working" at all! It is tested!) Approximately the same words are referred to the initial points **B**, although here it is no longer so critical.

(... By the way, here you can ask a million questions: "What will happen if we take not a square, but a rectangle", "what will happen if we take the mask with the symmetries of a rectangle, not a square", "what will be ... "? But to begin answering them and giving schedules - would mean to go the "wrong way". Our urgent task is to drive the behemoth in a matchbox. And so we shouldn't distract.

We'll give just two figures. Let us ask ourselves: how the "median" of the value $T_{1/2}$ - a number that for 50% of the tests under these conditions will be less than his, and for 50% it will be bigger - depends on the size of the site? The answer is: for the mask of the square 9x9, on the segment with sizes 30 - 80, with an increase in the size of the square for 10, it increases approximately in 2.2 times. The dependence is approximately linear.

Let`s wonder what will be with an increase of the number of our original cells **B**. The answer is: with an increase of the number of the cells, the situation in the ideological sense, becomes much simpler! The question "will the soliton return in a small number of steps or no" depends on: whether will be the SuperRiver formed in the first 100 - 200 steps? If it is formed - then everything is fine! It will surely - with a guarantee! - "clean" everything, and the soliton will close in less than ~ 20000 steps. If the answer is "no" – then the time of returning will be much, much more. Incidentally, this fact allows us to find easily absolutely magnificent examples of the "returns"! We can take a very large, for a home computer, the area size: 120 x 120. VERY big for it the value of number of initial points: 30. And then, take and sort out random Automata. Each time we should literally look through the first few hundred steps, if the SuperRiver was formed or not? If - "no", we should go to the next set. But if it is suddenly formed, then the soliton will surely close within the value of ~ 100,000 steps.

For our "good masks" the value of the number of points for which in half cases the SuperRiver is formed, and in half – isn`t formed, is ~ 16. That is, if in the table there were the median of appropriate half-cycles, approximately to the value of 16, there would stand a number ~ 15,000, and after – it would jump very abruptly.

We believe that for the first acquaintance it is enough! These examples will be enough for us till the end of the article.

Briefly we`ll give the last thing that is shown on the figure 5. Green highlighted solitons are those, which will be described in Chapter 6. In the upper right corner the characteristic numbers of these solitons are shown. See Chapter 7).

## 4 How to examine DCS. Filters

Just look at the soliton - is not correct. We should look at it through the filter. And our goal is to make such filters that would work "in two directions": in the direct (in the direction of time), and vice versa. And to portray it **between** two successive frames. For this we take six consecutive frames of motion of our soliton, as it is shown in the figure 4. The frames 0, 1. 2, 3, 4, 5. Just in the middle - between the frames 2 and 3 - and we shall represent a "filter". Our picture of all these six frames.

1). Our first filter

$$A_F(t, x, y)$$

– the filter of the number of cells A in our six frames t-3, t-2, t-1, t, t+1, t+2 which have the coordinates x and y. It may have 7 values, from 0 to 6. Let`s note an obvious formula (see figure 4).

$$A_F(t, x, y) = A_F(2T_{1/2} - t, x, y)$$

for any t, x, y.

We know that during steady motion of a soliton, the cells change, in general, **CBCBAA** or **CBACBA**, with different phases, and this means that in most cells for a variety of frames $A_F = 2$. (By the way, these two cases, sort out all the possibilities for exactly two cells **A** in six frames). So let's paint the first case (**CBCBAA**; Bank cells) gray, and the second one (**CBACBA**; River cells) green. The other colors are shown in figure. However, except for the first few frames, the number of $A_F$ could basically be equal only to 1 or 3. Well, maybe (rarely), to zero or to four, when some river, oscillating, passes through 0.

At once we should note very beautiful objects, which in the original, not filtered image were not visible. We called them NullRivers. Typically, they enter in gearing with the ordinary Rivers or with the others "Null". But we can easily find the pure one. In the figure 7 there are two "pure" at once.

NullRiver - is, indeed, the river with the length of zero. This is a single point of another color standing on the bank at a sufficient distance from the nearest River. And although the location of the nearest cells around it is different, the NullRiver on its own looks at our filter wonderfully without a single defect. (We should note that its period is equal to 12, and that their position is not directly connected IN NO WAY with the initial cells **B**! They are formed in absolutely random places). We'll talk about this in Chapter 9.

2). Our second filter

$$B_F^{\,i}(t, j, x, y)$$

This filter puts the color **on the boundaries** (boundary) **between the cells** and fixes the "changes" at the borders. It depends on the same index "i", which can have three values: 0, 1, 2. ("j" has two indexes, and shows what kind of boundary is it - the horizontal or the vertical; x and y - as before, are the coordinates of the cell).

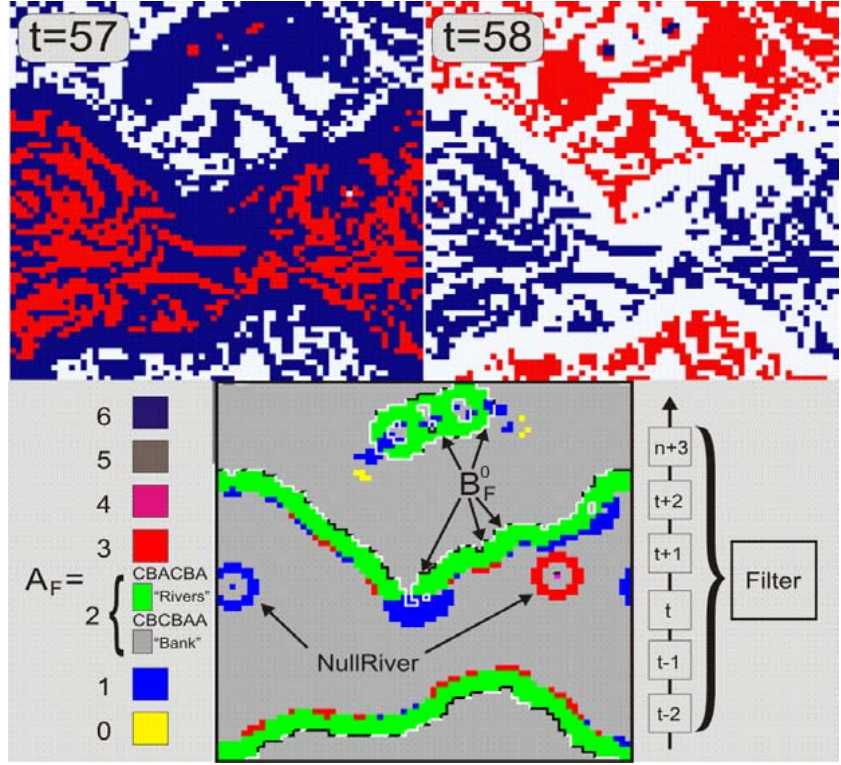

Fig.7. The filters $A_F$ и $B_F^0$ drawn between the frames 57 и 58 of a soliton 1(e). The value of half-cycle for it is 2526. So for the filter $A_F$ (the coloring of the cells themselves) drawn between the frames (2*2526-57, 2*2526-57+1) = (5001,5002) will look exactly the same and the coloring of the boundary $B_F^0$ will change to «the opposite».

If the value of i = 0 - we take "the changes" between the frames t and t +1. If i = 1 "the changes" between the frames t-1 and t +2. If i = 2 'the changes' between the frames t-2, and t +3. What exactly do we mean by the word "changes"? Let`s imagine that on the all the boundaries of cells there are put numbers "0" or "1". "0" - if the boundary separates the cells of the same color and "1" - if their color is different. Let`s take, for example, two central frames t and t +1. If there was at the border (at time t) a "0" and became (at time t +1) a "0" or it was "1" and became "1" – let`s then set for this boundary $B_F^0 = 0$, and we will not paint it. If there was a "0" and became "1", let`s then set for this boundary a number $B_F^0 = -1$ and paint the border in white. If there was a "1" and became "0", let`s then set for this boundary a number $B_F^0 = 1$, and paint it in black. Let`s note the obvious formula:

$$B_F^i(t, j, x, y) = -B_F^i(2T_{1/2} - t, j, x, y)$$

If we look at the figure 7 and wonder how will our filters look like for the same frames, but **after** a half-cycle (see fig. 4), the coloring of the cells will not change, and the coloring of the boundaries **will change to the opposite**: the black boundaries will become white, and the white will become black.

For the Filter $B_F^i$ two simple formulae which will be true for any reversible CA can be proved.

$$B_F^1(t, j, x, y) = B_F^0(t-1, j, x, y) + B_F^0(t, j, x, y) + B_F^0(t+1, j, x, y)$$

$$\sum_{t=t_1..t_2} B_F^0(t, j, x, y) = k \Rightarrow k \in \{-1; 0; 1\}$$

But, in general, the filter $B_F^i$ disappoints. Even seemingly true statement about the closure of "black and white" lines in fact turns out not to be true. For the mask (e) the violations also exist, although they are rare. For the mask (j) - they are ubiquitous. It seems that the only thing for which they can be useful - is to determine the "direction of movement" of the Rivers and SuperRivers in space, what we are going to talk about in the next chapter.

# 5 Rivers and SuperRivers through filters

3). Let`s introduce a third filter

$$C_F^{\,i}(t,x,y)$$

it also depends on the three indices, as a filter $B_F$, but paints in turn, not boundaries, but the cells, as a filter $A_F$. (C-coinsedance; the filter of "matches **A**").

Let`s suppose that we are building this filter for a zero index, that is, to the frames t and t +1. We`ll mark those cells that are in a condition **A** on these frames as a 1, and those which are not as a 0. And we`ll choose the function "Exclusive OR" between them. The result is our filter $C_F^{\,0}$. The filters $C_F^{\,1}$ and $C_F^{\,2}$ are constructed similarly. And if it is equal to 1, we will denote this filter with a small black cross in the center of the already colored by the $A_F$ filter cell. For it, as for the filter $A_F$, obviously, the same formula is true:

$$C_F^{\,i}(t,x,y) = C_F^{\,i}(2T_{1/2} - t, x, y)$$

Let`s examine how look all three filters for the half of the separate SuperRiver. See fig. 8.
They look very typical! Black and white curve of the filter $B_F^{\,0}$ alternately passes from one side to another and then disappears for one frame. The polylines of the filter $B_F^{\,1}$ border the whole river. Near the river, touching it, there are only "red" and "blue" cells. (We`ll recall that the "blue" cell is one cells **A** in six frames, "red" - cell is three). There is no others! The filter $C_F^{\,1}$ also looks very beautiful. It fills alternately one half of the torus relatively to the SuperRiver, then a second half, then both of them...and then all repeats! Obviously, this is the way a pulsating River looks. During pulsation the phase changes and the rhythm is got out.

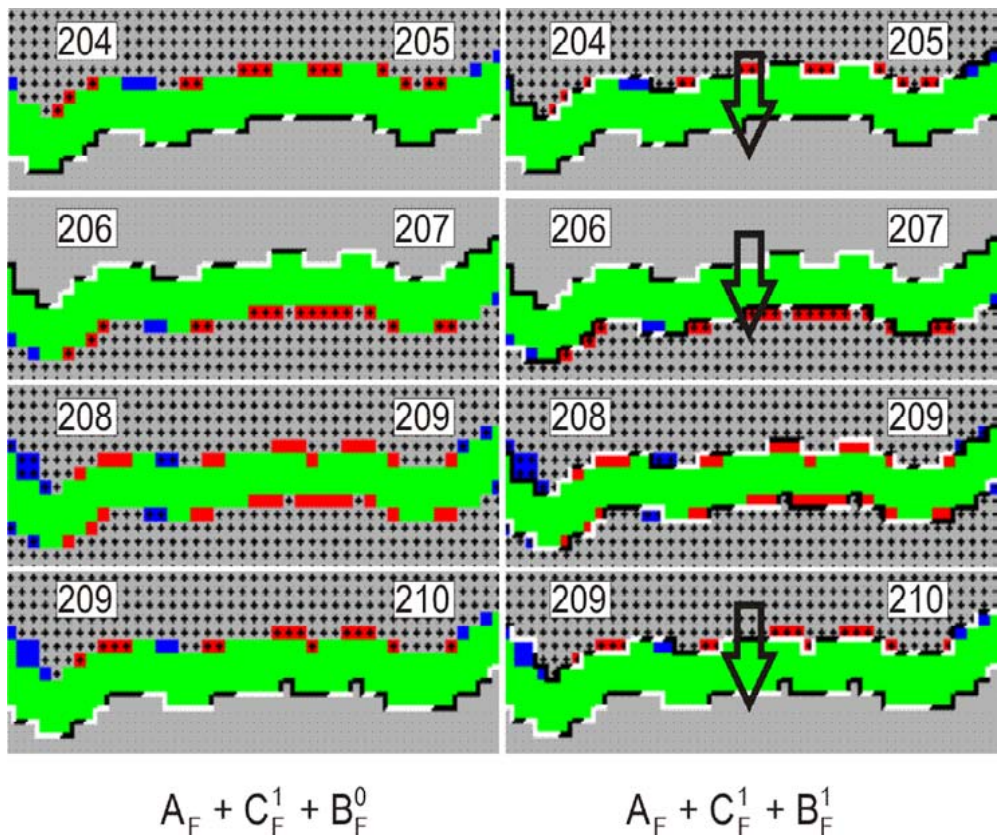

Fig. 8. Four frames of the filters of the movement of our soliton 2(e). The arrows indicate the "direction" of our rivers in the area. The arrows of the second half of the SuperRiver are directed toward them. All the arrows point to the location of future pulsation.

Let`s suppose that our SuperRiver is ready to pulsate. How with the help of the filters, looking only for a few frames, establish in which half of the torus it is going to happen? Curiously, this seems to need all three of our filters at once.

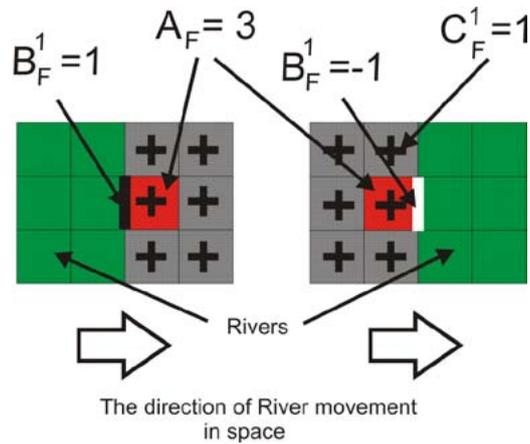

Fig.9. The determination of the movement direction of the half of the SuperRiver with the use of our filters.

It is easy to see that for one half of the SuperRiver these filters, which are - $A_F$, $B_F^1$, $C_F^1$ - can occur only in the configuration shown in the figure 9. This makes it possible to determine the direction of its movement. In our case, the arrows will point to the location of pulsation. When the SuperRiver passes the pulsation, the arrows will point to the contrary - "to the outside".

Let`s wonder if it is possible to guess from the same number of frames of the SuperRiver, in which half (before the half-cycle or after) we are?

The answer is no! The SuperRivers outwardly look exactly the same! And what can we say? Very little! We can only say that **exactly in this half of the torus**, sooner or later, "something interesting" will happen! The River will either close quietly, or vice versa, will produce the new Rivers ... then new ones ... then "anti-explosions", the Start Point, an "explosion" ... and so on. But we can not say what exactly! The author has an intuitive assumption that it is fundamentally impossible.

Actually, the filter $B_F^1 + C_F^1$ from the most elementary steps impress very much: with the closure of the filter $B_F$, and with the fact that, $B_F^1$ mainly goes in the boundaries of the areas with $C_F^1=0$ и $C_F^1=1$...but only for masks (a), (b) , (c)! For the mask (e) and especially (j) there are different disorders. But we do know that all these masks return equally well, and this means that in our consideration these filters are useless!

## 6 Key DCS graphics. "Local" time reversal

Obviously, if there are numbers $N_C(0)$, $N_B(0)$ and $N_A(0)$ (initial number of cells **C, B, A** in the automaton), and the graph $N_C(t)$ (number of cells **C** after t steps), we can immediately build the other graphics: $N_B(t)$, $N_A(t)$. Therefore, we will consider only the graph $N_C(t)$.

Later ... Approximately by the tenth run our River Systems with Different Banks is forming well. And from the assumption that any Bank has approximately equal number of the cells in 2 colors follows that this graph ($N_C(t)$) is roughly continuous with a period of three. Therefore, we will consider the $N_C(t)$ with three phases. The green line will denote the phase 0 - $N_C(t/3)$. The blue one - phase 1, $N_C(t/3+1)$. And the red one - phase 2, $N_C(t/3+2)$. (t/3 is the integral part of the division of t to 3).

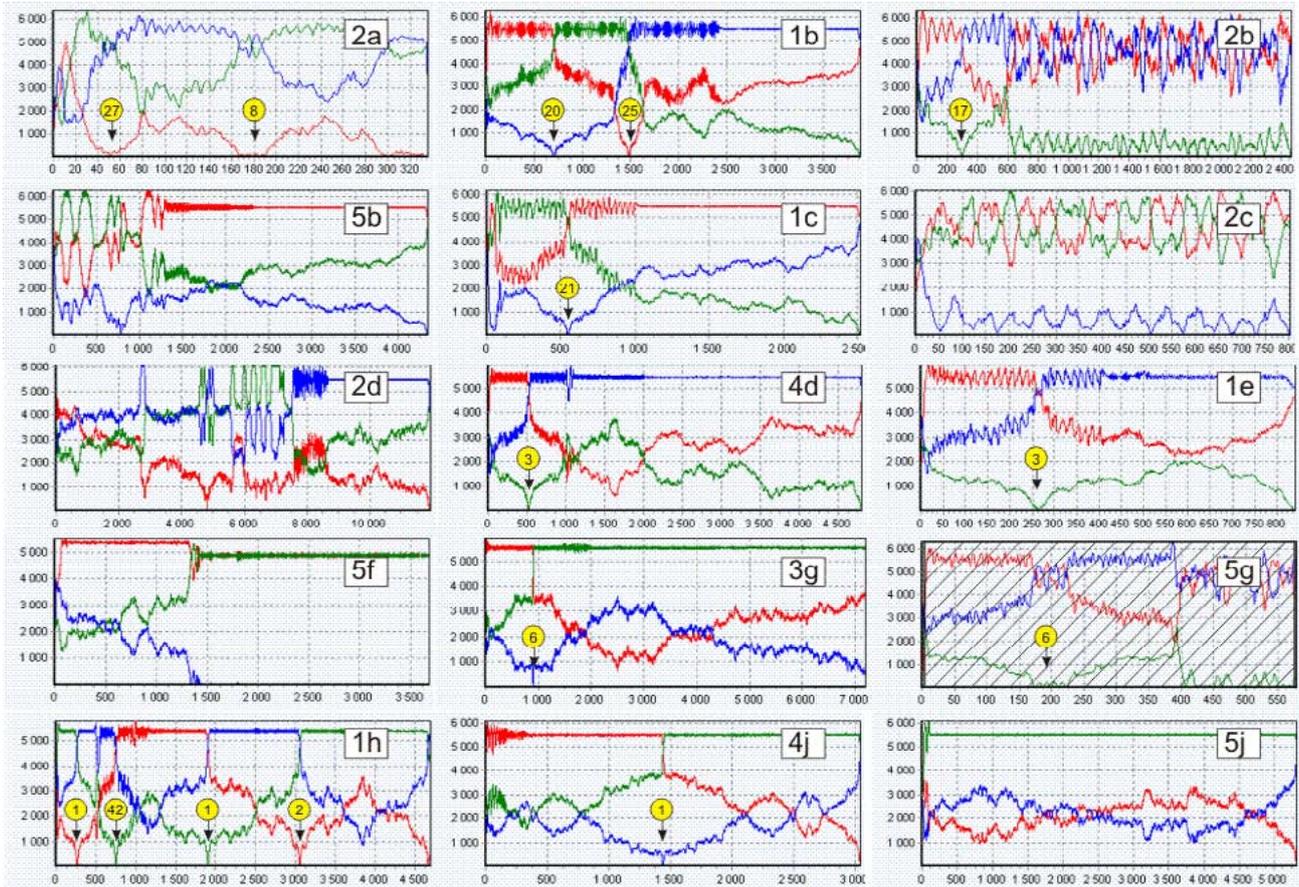

Fig. 10. Twelve of the fifty graphs $N_C(t)$. All of them returned in 100,000 steps with the exception of case 5(g). Yellow circles mark the moments of **local** time reversal. A number inside them is the number of cells **C** at this time. (That is, at the very end of the all cases, except for the 5(g), there is a "yellow circle" with the number 0. This is the moment of **real** time reversal).

The figure 10 shows the 15 graphs $N_C(t)$ of the movement of our solitons. All the cases returned, except for the 5(g).

For the cases which returned it is easy to understand how will look our graphs after the point of half-cycle. Let`s take as an example the case 1(c). There the green phase turned its $N_C$ into zero, and hence it will reflect strictly symmetrically relatively to the point of half-cycle. The remaining two phases will also reflect strictly, but the red line will go into blue and blue into red.

Let`s consider the points marked with yellow circles on the graphs. It is evident that this is "almost" time reversal, or reversal in a large area of our torus. We will call them **local** time reversal. (The number in the circle itself is the number of $N_C$ at this point). There are many mysterious things here. Let`s take, for example, the case 1(e). There is present the pulsing SuperRiver and somewhere near there is a large cluster of Rivers and NullRivers. And ... at the very moment of pulsation, when the number of $N_C$ in the area of SuperRiver reduces to one (!)... at the same time the cluster (for some reason?) reduces its value of $N_C$ to a record of two (!). The Automaton has passed the point of "local reversal" and "almost" went back in time. The number of cells **C** increases rapidly to several hundred! ..

But having come to its virtual "start", it would no longer be like before. In case 1(e) the automaton has worked for some time and closed successfully. The case 4(j) is interesting. The automaton came to a virtual start and shut down literally immediately. In the case 5(g) the automaton goes back in time almost to the time t = 0 ... and there is a "disaster"! The automaton by the explosion is divided into a very large amount of NullRivers and 100,000 steps were not enough for it to close.

But, generally, a variety of cases is amazing!

Why did the cases 5(f), 2(b) and 2(c) close?..It looks very incomprehensibly the "war" between two SuperRivers (the case 2(d ))?.. And how do you like the sequence {1, 42 (??), 1, 2} in the yellow circles of the case 1(h)?..

## 7  The Main Conservation Law of the DCS

Let`s go to the positive moments… Let`s ask ourselves: if it is «soliton», so where is the «infinite number» of the conservation laws?

We`ll find them easily!..

Thus, we`ll present our **Main Conservation Laws (MCL) of our soliton**:

**For any soliton, for any x and y**

$$\frac{1}{2}(A_F(0,x,y)-2) + \sum_{t=1..T_{1/2}-1}(A_F(t,x,y)-2) + \frac{1}{2}(A_F(T_{1/2},x,y)-2) = 4\lambda$$

**where $\lambda$ - is whole number equal to the constanta!**

Let's discuss it in details...If we take $A_F$ at the moment of time t = 0, it is easy to see that it is even in the all cells, since the frames -3, -2, -1 are the transliterations of the frames 0, 1, 2. For the same reason, the number $A_F$ at the moment of time t =$T_{1/2}$ is even too. Let`s subtract 2 from them, divide in half and add. And to the whole number, that we got as a result, we add a full-fledged $A_F$-2 throughout the motion from t =1 to t = $T_{1/2}$-1.

As a result, in all cells we`ll obtain **the same number multiple of 4**. We`ll divide it by 4 and call as a characteristic number of this soliton. By analogy we`ll denote it with the number $\lambda$. (It`s they which are given in the table in the figure 5, in the upper right corner).

The number of conservation laws is not "infinitely large" as in the usual soliton, but it is also very considerable. Their amount is as big as the amount of the cells in our area, namely 70*70 = 4900!

The law mentioned above is tested… maybe for 100 very different solitons! And the "right" ones (described in the article), and the "wrong" ones. (Our conditions may be changed a little – for instance, no to close one side in the torus, or to go to the site 70 x 71 - and the soliton will return again. Sometimes. So: the Main Conservation Law is still immutable!) These Laws are as if cementing our narrative. It becomes clear that we are talking about the Phenomenon, which is uniform in spirit. The author doesn`t know anything about the nature of these Laws and the meaning of $\lambda$, and he considers it inappropriate to talk on this topic.

The behavior of the function F from **t** is very interesting:

$$F(t,x,y) = \frac{1}{2}(A_F(0,x,y)-2) + \sum_{q=1..t}(A_F(q,x,y)-2)$$

.

First, it dramatically swells out on X and Y, and then, as if cools. But what is interesting is that almost at the all points *x, y* the function *F* on the whole movement - stays even. And always only in some points - it is odd. (Then it will be divisible by 4, and besides, it will be the same at all the points after the addition of the last "half". At the last step only a few points are added with $A_F - 2 = -2$ (can be zero points). And a few points with $A_F - 2 = 2$ (maybe, too, zero points). But, apparently, both of them can not be zero.)

## 8  The main integral of the DCS

The conservation laws lead to the concept of the Main Integral of the DCS - function - *S*. The values of all the sums values ($A_F$- 2) **for all *x* and *y*** from $t_0$ to *t*. (And if $t_0$=0 it is necessary to take the half of this sum. If t=$T_{1/2}$ then it`s necessary to do the same).

$$S(t_0,t) = \sum_{q=t_0..t}\sum_{x,y}(A_F(q,x,y)-2)$$

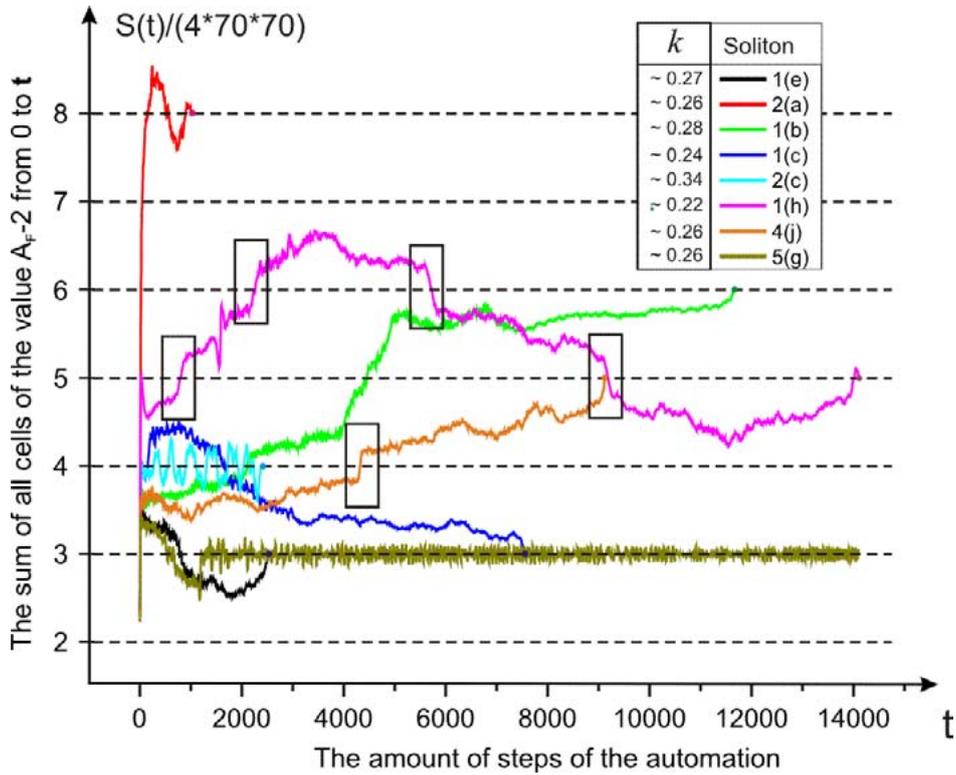

Fig.11. The behavior of the function *S(0,t)=S(t)* for eight solitons. Only one of them – the variant 5(g) didn`t return.

The figure 11 shows the function S(t)= S(0.t) for eight solitons.
In this figure it is perfectly clear that our "local time reversal" (some of which are outlined with rectangle) neither are a false artifact nor the "aberration of view". It is passage by the function *S* «not its λ». Well, maybe "its", but "not at this time '(soliton 1(h)). (The case 2(c) is special!) It is seen that the solitons prefer to do it at speed, and it seems so, that with the same derivative. Again, it seems that the variant 5(g) did not succeed in doing it.

It`s interesting that function S(t) may come both from above and below to the Point of half-cycle.
And now, the reason why we named our function S(t) – the Main Integral...

## 8 The main «symmetry» of the DCS

And our final observation...It concerns the connection of the function S and $N_C$. We called it the main "symmetry" of the DCS. (The word is in quotes because it is not quite correct, it is"fuzzy". But despite this, it is very, very "powerful" and therefore – the Main one). We are not denying the pleasure to tell you how we came to it...

When looking for a long time all over the figure 11, a strange feeling of "deja vu" appears. Have we seen it somewhere already?..Well, of course! On the figure 10. Moreover, we **have seen exactly these graphs**. But with a little explanation...

Let's go to the other end, and start from the Point of half-cycle. And at once we`ll write down our formula for symmetry.

$$M(t) \approx M_0 + kS^{\pm}(t) = M_0 + k\,\text{sgn}(M) \sum_{q=0..t} \sum_{x,y} (A_F(q,x,y) - 2)$$

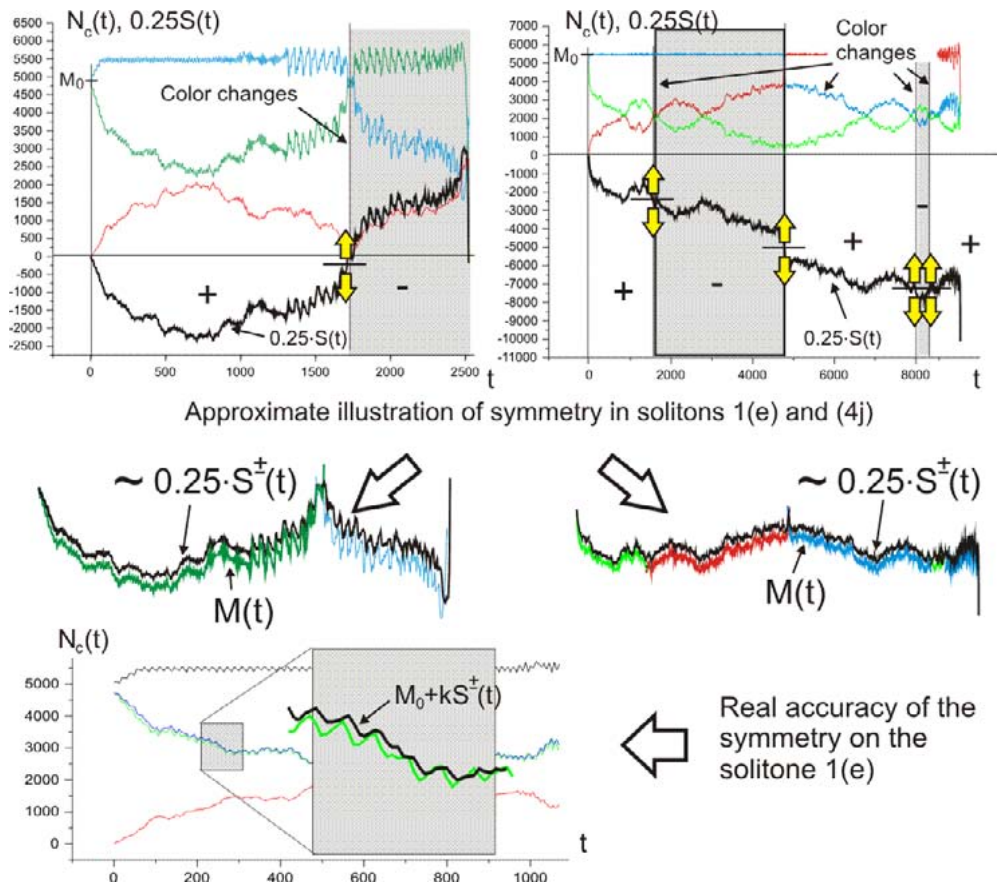

Fig.12. The illustration of the way how we came to the main «symmetry». (Everything is explained in the text).

And now we`ll leisurely explain everything shown on the figure 12.

Let`s arrange the graphics. To arrange the graphics $N_C$ and S on t it is just needed to reflect all of them relative to the vertical line. So let's do it. (We are not concerned that we may get the wrong colors!)

We should select *k* for the graph *S*. Let`s suppose that we have guessed and have chosen k = 0.25. (If we have not guessed, it is easy to adjust it to about exact). And we can see at once that the graph S(t) suspiciously exactly resembles...the "median" graph of our graphics $N_C$! The "median" graph - M(t) in the formula mentioned above – is the **average graph** of three; it is easy to see that it is also always about continuous.

But the second half of the black graph S, for some reason does not coincide with the median. It would be good to reflect it relative to the horizontal line at the point A. And then we see that at this very moment – at the point A - the colors on the median changed. It was green, and became blue! And we understand that the solution is found ... It is just necessary to introduce a signature - sgn(M) – for our Main Integral and change it to the opposite, when the median color changes. It's very simple. We check the same principle at the soliton 4(j). Here we have not only one, but 4 changes of color. We see that here everything also coincides.

Relative to the value $M_0$... It may be equal to the number of white cells in the picture of half-cycle, may be equal to the number of blue ones. Compare the graphics 1(b) and 1(c) from the figure 10. In the case of soliton 1(b) after a point of half-cycle the median graph will become red (phase = 2), and in the case 1(c) - blue (phase = 1). The initial signature depends on how the Main Integral function comes to the point of half-cycle: from the the top or from the bottom. (Fig. 11).

But if everything is chosen correctly, then on the figure 12 we see an "almost perfect" coincidence, the functions M(t) and S plus-minus from t. Then the graphics will slowly begin to diverge, and it is difficult to say what is the reason for this? The fact that we have chosen a little bit wrong k, or...The fact that there is an obvious drawback of the scheme. Every phase is average at the moment, we can determine the average once every three frames, and the total amount should be taken every frame! And we can show that it is important! The amount graph, quite a bit, but jumps at the moment of color change. And over time, it will "jump" far enough. (See fig. 13). But again you can "put" one graph to another, and continue ... Again, at first everything is perfect ... Curiously, that *k* does not change throughout the movement! This new constant is unique for each soliton. We have shown them for our seven solitons in the figure 11.

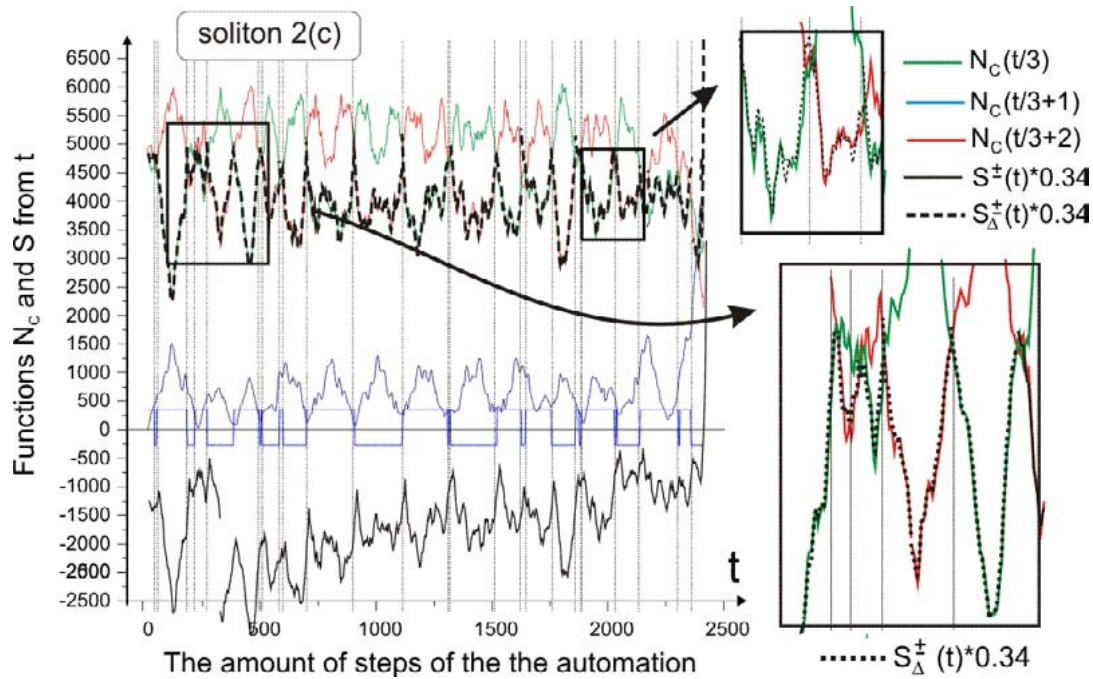

Fig.13. The illustration of the Main «symmetry» on a «difficult case» - soliton 2(c). With every changing of the color the graph S put on the median one. The permanency of the value k is well seen.

So we'll go all the soliton from the end to almost the very beginning. But the more we approach the current of "anti-explosion," the stronger will be "jumping", and at the very Start Point the real inconsistencies will appear... But our "symmetry," almost to the point X, will try to "save its blushes"! (Let`s wonder whether this symmetry will help to answer the old question: "In what part of the cycle are we - up to halfcycle, or after?" Of course not! In the other direction, all the signatures just will change the sign and everything will go as before)

Let`s have another look at the strange nature of our Main "symmetry." Every time it turns out that exists an own "leading" phase?! At the moment this is median phase. And it is equal to the total integral of a strange function? Again, nothing is clear!

## 9  Soliton in the other dimensions

Suddenly, the output in three dimensions greatly "clarifies" the situation! (It's amazing! We absolutely do not understand what do our "conservation laws", "lambda", "symmetry" mean... We do not even approximate idea how to prove the same MCL. But "from a philosophical point of view" everything already seems "to be clear for us"! Yes, it often happens!)

We reviewed our soliton in a cube with size 11 x 11 x 11 (cell number is 1331), locked all the opposite sides in it. For the mask we took a cube 3x3 cube without apices. (In total, in the mask there is 18 cells). The number of initial points **B** was taken as 8, 12, 16. For each of the variants we took fifty cases (totally 3*50 = 150 solitons). As the threshold «came – didn`t come» we took the same that was: 100,000 steps.

Totally during this time approximately 90% of all variants of the number of the cells **B** managed to close. For the case with 8 initial cells the median of the value $T_{1/2}$ was 8600. For 12 it was 9300. For 16 it was 22000.

We see that the DCS exists in the three dimensions and feels well there. By the way, better than in two dimensions! (Even partial no-closure of the sides of the cube is not fatal. Without closing of the four sides of the cube – about 35% still return, of the two sides ~ 65%.) Obviously, for our DCS – the more dimensions are the better!

Naturally, there are MCL. In 32 of 50 cases for the soliton with n = 12 and having a $T_{1/2}$ less than 20,000, "lambda" was distributed as follows. 19 - lambda is equal to 2, 11 – 3, 1 – 4, 1 - 5. All the cubes which are not closed into torus have the same clear "lambda".

But unlike the case of two dimensions, in the movement of 3-dimensional soliton immediately reveal a clear phase with a very small number of **C** cells. (Later, it may also change, but usually this does not happen). And since Nc is already close to zero, the Automaton immediately tried to organize a «local circulation.» Up to 5 (!) explicit areas of local circulation in some cases. Moreover, the local circulations can last quite a long time. Some of the them with the duration of ~ 10,000 steps! But Rivers and SuperRivers are not formed any longer in three dimensions. So what happens? The thing is that the whole issue of return assume ... NullRivers! And they cope quite well with it!

Let`s look at them closer. At the same time in two and three dimensions.

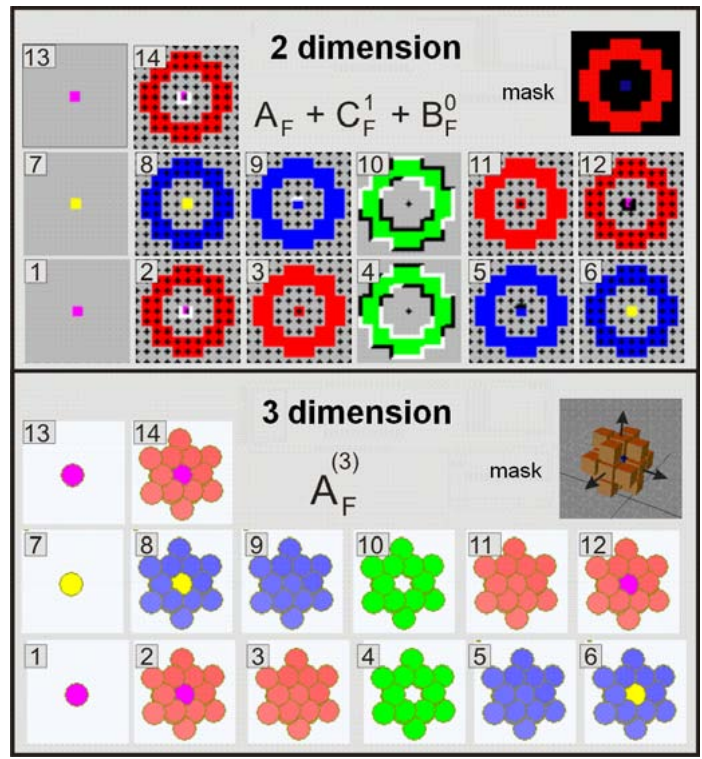

Fig. 14. NullRivers for two dimensional and three dimensional masks. (We would like to recall that the colors Yellow, Blue, Red, Purple correspond to the $A_F$ equal to 0, 1, 3, 4. The Green color is River $A_F = 2$) In the case of two dimensions all the filters are given. In the case of three dimensions there is only a filter $A_F$. The Bank in three dimensions is considered to be transparent.

The complete similarity of the two NullRivers is visible. Obviously, in any dimensions and in any masks the NullRivers look the same. (We would like to recall one more time, what the River of Zero Length (NullRiver) means. It is a single point of a different color, standing deep on the Bank. The Bank itself has a period 6. And when there is a point of another color, their common period doubles).

But the other thing is interesting! It is clearly seen that the NullRiver – *is organized*! It stands among a random pile of other cells, but by itself - it is perfect and without a single defect. There is also a random outlining of its mask. This is its *code*! The outlining of the center cell – is also its *code*! In the frames 2 and 9 there can be only *white* outlining. And in sum, it should close in the square. The outlining of the center cell in the frame 12 - is opposite to the frame outlining 2. The outlining in the frame 5 – is opposite to the frame outlining 9. ("Can white outlining of the frames 2 and 9 have "diagonal symmetry"? Let`s say, a corner at the bottom left and at the top right?" The author especially were looking for it for an hour on the computer for a great variety of masks – and he didn`t find! It seems that there is no... such, but I will not lie!) And that's the way how **every** (!) NullRiver is organized.

Let us ponder over what is going on! We take any number of dimensions. Any mask. (Well, almost any!) Switch the Automaton on. Blast! The Universe is born. It would seem, that any traces of the starting points **B** should be completely washed away. The Automaton continues to work... And suddenly to our surprise we found out that everywhere **the same** ... usually formed in the strange linkages but exactly **the same** objects! Which start focusing on their main task - a joint return to the original Universe. (And for completing this task they only need to organize themselves! Or how otherwise? They ALL TOGETHER have not to fall into a Cycle! The one can`t avoid getting into the Giant Cycle!) Here is an alone NullRiver which is interacting intensively with its neighbors! OH, IT IS NOT A SIMPLY TASK – to come back from the plus infinity in 7000 steps!..

And, in general, do not you think, looking at the figure 14, you're looking at the Perfection?..

And I think so! And I sometimes think that the essence is in that. That NullRiver, recognizing itself as a Magic Ideal of Math, naturally sets itself the EXTRA tasks. And what EXTRA could have any creature? Of course, search of Immortality, *as they understand it*! The NullRivers understand it, how not to fall and not dissolve completely in the Giant Cycle!

And just two dimensions – it is a singular case. The amount of measurements is too small! And the NullRivers simply can not wriggle in order to "realize their dream". That`s why they have "to change the strategy". To cooperate with SuperRivers, with common ones... and finally reach the goal! Even in three dimensions, nothing of this is necessary. They directly reach their aims!

## 10  Conclusion

The name that we have chosen for our phenomenon may cause some criticism: "soliton". "Soliton" – means solitary. In the sense of "solitary wave." "Where is the wave here?"

One can find a few similarities between these two phenomena. But we will say, from our point of view, about the main one.

For us "Soliton" is a symbol of complete INCOMPREHENSIBILITY, "how could it be". Even with the usual soliton not everything is clear, despite the tremendous efforts expended. And our cellular soliton, when you meet it, leaves exactly the same stunning impression...

"The time that is always running at the same time forth and back..." "The space that completely transforms into the time, and then, on the contrary, back into space..." (This is how we can interpret our Conservation Laws). "Some absolutely not understandable constants..." "It is absolutely unclear symmetry..."

And it turned out that the show is run by strange Ideal Objects...

Which, in awareness of their own Beauty and Perfection, reached the point where they dared to defy the Gods themselves... and so, oddly enough, but the Gods had to make concessions!

«WELCOME TO THE WORLD OF CELLULAR SOLITON!»

The world is SO "abstract"... that, in comparison with it, any Wonderland or Through the Looking Glass, from the "Alice's Adventures" - is simply a top of old English decency and common sense!

And what is the most surprising is that all this is just one step away, literally, under our noses!

And all we need ... is to turn our home computer on and do this step.